\documentclass[a4paper, 12pt, openbib ]{article}
\usepackage{times}
\usepackage{pifont}
\usepackage{amsmath}
\usepackage{amssymb}
\usepackage{amsfonts}
\usepackage{graphicx}
\usepackage{floatflt}
\usepackage{geometry}
\usepackage{layout}
\usepackage{enumerate}
\usepackage{bm}
\usepackage{tikz}
\usetikzlibrary{patterns}
\usepackage{amsthm}
\usepackage{thmtools}
\def\tfract#1/#2{{\textstyle{\raise0.8pt\hbox{$\scriptstyle#1$}\over%
\hbox{\lower0.8pt\hbox{$\scriptstyle#2$}}}}}
\def\mezzo{\tfract 1/2 }

\def\gmezzi{\tfract g/2 }
\def\gsesti{\tfract g/6 }

\def\sesto{\tfract 1/6 }

\def\radi2k{\tfract 1/{\sqrt {2k}} }
\def\der{\partial }
\def\cvd{\vbox{\hrule \hbox to 9 pt {\vrule height 9 pt \hfil \vrule} \hrule}}
\def\downnormalfill{$\,\,\vrule depth4pt width0.4pt
\leaders\vrule depth 0pt height0.4pt\hfill\vrule depth4pt width0.4pt\,\,$}
\def\WT#1{\mathop{\vbox{\ialign{##\crcr\noalign{\kern3pt}
      \downnormalfill\crcr\noalign{\kern0.8pt\nointerlineskip}
      $\hfil\displaystyle{#1}\hfil$\crcr}}}\limits}

\def\be{\begin{equation}}
\def\ee{\end{equation}}
\def\bes{\begin{equation*}}
\def\ees{\end{equation*}}
\def\bea{\begin{eqnarray}}
\def\eea{\end{eqnarray}}
\def\beas{\begin{eqnarray*}}
\def\eeas{\end{eqnarray*}}
\def\ba{\begin{array}{rcl}}
\def\ea{\end{array}}
\def\der{\partial}
\numberwithin{equation}{section}
\def\go{\leavevmode \raise.3ex\hbox{$\scriptscriptstyle \langle\!\langle\!  $}%
~\ignorespaces}
\def\gf{\relax \ifhmode \unskip~\else \leavevmode \fi \raise.3ex\hbox{$\! \scriptscriptstyle\rangle\!\rangle\, $}}
\usepackage{amsmath,amssymb,amsfonts}	
\usepackage{slashed}            
\usepackage{bbm}                
\usepackage{xspace}				

\usepackage{tikz}
\usetikzlibrary{patterns, arrows, shapes}
\usetikzlibrary{trees}
\usetikzlibrary{matrix, arrows} 				
\usetikzlibrary{positioning}				
\usetikzlibrary{calc,through}				
\usetikzlibrary{decorations.pathreplacing}  
\usepackage{pgffor}							

\usetikzlibrary{decorations.pathmorphing}	
\usetikzlibrary{decorations.markings}
\tikzset{
    vector/.style={decorate, decoration={snake}, draw},
	provector/.style={decorate, decoration={snake,amplitude=2.5pt}, draw},
	antivector/.style={decorate, decoration={snake,amplitude=-2.5pt}, draw},
    fermion/.style={draw=black, postaction={decorate},
        decoration={markings,mark=at position .55 with {\arrow[draw=black]{>}}}},
    fermionbar/.style={draw=black, postaction={decorate},
        decoration={markings,mark=at position .55 with {\arrow[draw=black]{<}}}},
    fermionnoarrow/.style={draw=black},
    gluon/.style={decorate, draw=black,
        decoration={coil,amplitude=4pt, segment length=5pt}},
    scalar/.style={dashed,draw=black, postaction={decorate},
        decoration={markings,mark=at position .55 with {\arrow[draw=black]{>}}}},
    scalarbar/.style={dashed,draw=black, postaction={decorate},
        decoration={markings,mark=at position .55 with {\arrow[draw=black]{<}}}},
    scalarnoarrow/.style={dashed,draw=black},
    electron/.style={draw=black, postaction={decorate},
        decoration={markings,mark=at position .55 with {\arrow[draw=black]{>}}}},
	bigvector/.style={decorate, decoration={snake,amplitude=4pt}, draw},
}

\tikzstyle{block} = [draw, rectangle, 
    minimum height=3em, minimum width=6em]

\title{
{\Large  \bf Gauge fixing and metric independence \\ 
in topological quantum theories}
{\vskip 0.6 truecm}} 

\author{{\large  Enore~Guadagnini$^{\, a}$, Federico~Rottoli$^{\, b}$ and Frank~Thuillier$^{\, c}$} \\  {\normalsize {~}} \\  
{\normalsize  $^{ a\, }$Dipartimento di Fisica {\it E. Fermi} dell'Universit\`a di Pisa, and INFN Sezione di Pisa,} \\ 
{\normalsize   Largo B. Pontecorvo  2, 56127 Pisa, Italy.}
 \\ {\normalsize $^{b\, }$SISSA - International School for Advanced Studies, via Bonomea 265,} \\ 
 {\normalsize and INFN Sezione di Trieste, Trieste, Italy. }\\ 
 {\normalsize $^{c\, }$LAPTh, Universit\'e Savoie Mont Blanc, CNRS, Chemin de Bellevue, BP 110,} \\ {\normalsize     F-74941 Annecy-le-Vieux Cedex, France.}
}

\date{}

\begin{document}

\maketitle 

\vskip 0.7 truecm 

\begin{abstract}

We consider topological gauge theories in three dimensions which are defined by metric independent lagrangians. It has been claimed that  the functional integration necessarily depends nontrivially on the gauge-fixing metric. We demonstrate that the partition function and the mean values of the gauge invariant observables do not really depend on the gauge-fixing metric.

 \end{abstract}

\vskip 1.2 truecm

\section{Introduction}

The three-dimensional quantum Chern-Simons theory \cite{WI} and the BF model \cite{HO,KR,MP,BT,GR} are gauge theories characterised by  metric independent lagrangians.  This implies that the observables defined in these theories are necessarily related with topological invariants. But in perturbation theory,  in order to define the vacuum expectation  values of the product of the gauge fields, one needs to introduce a gauge-fixing procedure. Usually, the gauge-fixing lagrangian term $ L_{\phi \pi}$ depends on the metric $g_{\mu \nu}$ that one can introduce in the 3-manifold $M$, where the theory is defined. The addition of $ L_{\phi \pi}$ to the original  lagrangian modifies the correlation functions, but it does not modify the expectation values of the gauge invariant functions of the fields.  Thus the observables computed in the gauge-fixed Chern-Simons  and BF theories do not depend on $g_{\mu \nu}$ and represent topological invariants. This has indeed been verified in several examples. 

It has been claimed  \cite{WI,WDG,AT} that the partition function $Z$ of the Chen-Simons theory  nontrivially depends on the choice of the gauge-fixing metric $g_{\mu \nu}$.  More precisely, it has been envisaged that $Z$ should depend on  the choice of the trivialisation of the tangent bundle associated with $( M, g_{\mu \nu} )$   through the presence of an overall  phase factor \cite{WDG,AT}
 \be
Z = e^{i 2 \pi d \, \sigma ( M, g_{\mu \nu}) / 8 } \, Z_0 \; , 
\label{1.1}
\ee
where $Z_0$ is metric independent. It has been argued that the function $\sigma ( M, g_{\mu \nu})$ nontrivially depends on the gauge-fixing metric \cite{AT} and can also be expressed \cite{AT} by means of the Chern-Simons functional computed with the Levi-Civita connection associated with the metric $g_{\mu \nu}$. Finally, it has  been claimed  that $\sigma ( M, g_{\mu \nu})$ does not depend on the coupling constant of the Chern-Simons theory.  

In the present article we shall investigate  the possibility that $Z$ necessarily depends on the gauge-fixing metric $g_{\mu \nu}$ in a nontrivial way.  We shall firstly concentrate on  the particular metric  dependence shown in equation (1.1); then we shall analyse a possible general metric dependence of the expectation values of the gauge invariant observables.   

When the Chern-Simons and BF theories are defined in topological nontrivial manifolds, with nontrivial homology group, because of the presence of zero modes one cannot follow the usual rules of the perturbative approach and of the standard gauge-fixing procedure.   Thus, in the present article 
we consider the topological Chern-Simons and BF theories in ${\mathbb R}^3$, where a canonical quantisation can be presented and a well defined  gauge fixing  procedure exists. We compute all the Feynman diagrams that could contribute to the phase factor $\sigma ( M, g_{\mu \nu})$ appearing in equation (\ref{1.1}).  We show that this overall phase factor is really not present. In our discussion, we recall some general arguments that confirm the validity of  the result of the explicit computations. 
Then we present a general argument showing that a generic dependence of the observables  on the gauge-fixing metric is excluded. 

In ordinary gauge theories of Yang-Mills type in four dimensions,   the gauge independence of  the on-shell S-matrix elements has been examined  by means of several methods \cite{N,KSJB,BLE,AFF,SIB,HHK,AGM,DCI}. 
However, these arguments cannot be trivially extended  in  topological  gauge theories because  a consistent on-shell formulation of these models has not been found yet. This is a fundamental aspect of topological quantum field theories. In order to clarify this issue, let us consider for instance the CS theory in ${\mathbb R}^3$. The equations of motion imply that the gauge curvature must vanish. Since flat connections in ${\mathbb R}^3$  necessarily belong to the trivial gauge orbit, there are no on-shell degrees of freedom and thus the standard on-shell formulation of the CS theory turns out to be trivial. Nevertheless, the vacuum expectation values $\langle W_L \rangle $ of the traces of the holonomies associated the oriented framed links $\{ L\}$ in ${\mathbb R}^3$ are not trivial. In facts, the mean values $\langle W_L \rangle $ take the form of generalised link polynomials and correspond to an infinite number of inequivalent nontrivial topological invariants for links. The manifold ${\mathbb R}^3$  admits a canonical $(2+1)$ decomposition as a product  ${\mathbb R}^2 \times {\mathbb R}$  and the set of CS observables in ${\mathbb R}^3$ is really  nontrivial, but a corresponding on-shell formulation of the theory does not exist.  

The issue of the gauge independence must then be formulated for the gauge invariant observables, and in this article we present an off-shell demonstration of  the gauge independence of observables in topological gauge theories. 
This argument has been considered in literature \cite{KAU,CUG,MAT,DAN,BRO,ABU,RES,CAIC,MTA,CHE,QUI}. The novel aspects of our proof are in order. We show that both the Chern-Simons and the BF theory admit a unified analysis, this was not fully recognised in previous papers. In order to deal with well defined expressions, we consider normalised partition functions and we explicitly compute the relevant one-loop Feynman diagrams. We shown that the ghost contributions ---that have been ignored in  previous articles--- play a crucial role in eliminating  the dependence of the partition function on the gauge-fixing metric. 
  Our  general argument of the  metric independence of the partition function and of the observables  is based on a generalisation of the Nielsen equation \cite{N} in which we consider the variation of the expectation values  with respect to a global parameter multiplying the whole set of gauge-fixing and ghosts terms.

\section{Quantum field theory fundamentals}

The nonabelian Chern-Simons  action in ${\mathbb R}^3$  is given by 
\be
S_{CS} [A] =  {k \over 4 \pi }  \int d^3 x \, \epsilon^{\mu \nu \rho}  \left \{ \mezzo A^a_\mu \der_\nu A^a_\rho
 - \sesto f^{abc }\, A^a_\mu A^b_\nu A^c_\rho  \right \} \; , 
 \label{2.1}
\ee
where the real parameter  $k$ denotes the coupling constant, the one-form components $ \{ A^a_\mu (x) dx^\mu \}$ locally represent  the gauge connection and $\{ f^{abc} \}$ designate  the structure constants of the Lie algebra of the simple gauge group $G$. Since the components of the connection are not gauge invariant,  in order to define a propagator for the  $ A^a_\mu (x)$ fields and determine the correlation functions of $ A^a_\mu (x)$, a gauge fixing procedure must be introduced.  

Note that, if one is interested on the mean values  of gauge invariant variables exclusively, in certain cases there is no need to fix the gauge. We shall return to this point  in {\rm Section~3}.

As it happens in the Fock-Schwinger gauge \cite{FO,SC},   certain gauge fixing constraints do not necessitate the introduction of a metric  \cite{GAL,FR}; in these cases, the partition function and the mean values of the observables  are automatically metric independent. Let us now consider the case in which a metric enters the  gauge-fixing method. 

 In agreement with the BRST procedure \cite{BRS,2S,TYU}, let us introduce the auxiliary field $M^a(x)$ and the ghosts and antighost  fields $c^a(x)$ and $\overline c^a (x)$. The BRST transformations take the form   
\bea
\delta A_\mu^a =\der_\mu c^a - f^{abc} A_\mu^b c^c  \; &,& \;  \delta c^a = \mezzo f^{abc} c^bc^c \; , \nonumber \\
 \delta {\overline c}^a = - M^a \; &,& \; 
 \delta M^a =0 \; .   
 \label{2.2}
\eea
In the Landau gauge, the   gauge fixing term reads 
\be
S_{\phi \pi } = \lambda \int d^3x \, \sqrt g \, g^{\mu \nu } \, \Bigl \{ \der_\mu M^a  A^a_\nu +  \der_\mu  {\overline c}^a  (\der_\nu c^a - f^{abc} A_\nu^b c^c )  
 \Bigr \} \; ,  
 \label{2.3}
\ee
in which $\lambda \not= 0 $ is a real parameter, and   $g_{\mu \nu}$ denotes a given euclidean metric in ${\mathbb R}^3$ that we call the gauge-fixing metric.    Note that $S_{CS}$ and $S_{\phi \pi }$ are separately invariant under BRST  transformations; thus $\lambda $ is a free parameter. 
 
 The total action $S_{TOT}$ is equal to $S_{TOT} = S_{CS} + S_{ \phi \pi }$ and the partition function $Z$ can be written in the form 
 \be
 Z = N^{-1} \int D ({\hbox{ fields }}) \, e^{iS_{TOT}} \; , 
 \label{2.4}
 \ee 
where $N$ represents a normalization factor     which needs to be introduced in order to specify the functional integration; we require that $N$ does not depend on $g_{\mu \nu}$. 
  Since the metric $g_{\mu \nu}$ enters  the definition (\ref{2.3}) of $S_{\phi \pi}$ though the the combination $ \sqrt g \, g^{\mu \nu } $,  it is convenient to add and subtract the identity $3\times 3$ matrix $\delta^{\mu \nu}$ and put
  \be
  \lambda \, \sqrt g \, g^{\mu \nu } = \delta^{\mu \nu} + h^{\mu \nu } \; , 
   \label{2.5}
\ee
 where
 \be
 h^{\mu \nu } =  \lambda \,  \sqrt g \, g^{\mu \nu } -  \delta^{\mu \nu} \; . 
 \label{2.6}
 \ee
 The total action $S_{TOT}$  can  then be written as 
 \be
S_{CS} + S_{\phi \pi} = S_{TOT} = S_0 + S_I \; , 
\label{2.7}
 \ee
 in which the ``free" action $S_0$ does not depend on the gauge-fixing metric $g_{\mu \nu}$,  
 \be
 S_0 = \int d^3 x  \left [   {k \over 8 \pi }   \epsilon^{\mu \nu \rho}   A^a_\mu \der_\nu A^a_\rho +  \delta^{\mu \nu }\der_\mu M^a  A^a_\nu +  \delta^{\mu \nu }  \der_\mu  {\overline c}^a  \der_\nu c^a  \right ] \; , 
 \label{2.8}
 \ee
whereas the ``interaction" action $S_I$ depends  on the metric $g_{\mu \nu} $ and takes the form 
\bea
S_I &=& \int d^3x \Bigl [  h^{\mu \nu} \der_\mu M^a A^a_\nu +  h^{\mu \nu } \der_\mu  {\overline c}^a  \der_\nu c^a \nonumber \\
&& \qquad -   {k \over 24 \pi } \epsilon^{\mu \nu \rho}  f^{abc }\, A^a_\mu A^b_\nu A^c_\rho - \lambda  \sqrt g \, g^{\mu \nu } f^{abc} \der_\mu  {\overline c}^a A_\nu^b c^c \Bigr ] \; . 
\label{2.9}
\eea
We can now specify the normalization of the functional integration (\ref{2.4}),  
\be
 Z = \frac{\int D ({\hbox{ fields }}) \, e^{i S_{TOT} }}{\int D ({\hbox{ fields }}) \, e^{i S_0 } } \; . 
 \label{2.10}
\ee
According to the standard rules of quantum field theory \cite{BO,IZ,PS}, expression (\ref{2.10}) corresponds  to the sum of the vacuum-to-vacuum Feynman diagrams which are constructed  by means of the interaction vertices contained in $S_I$ of equation (\ref{2.9}) and the field propagators determined by $S_0$ of equation (\ref{2.8}). 
For the components of the fields propagator we find 
\bea
\WT{A^a_\mu (x) A}\! \null^b_\nu(y) &=& \frac{4 \pi}{k} \delta^{ab} \int {d^3p \over (2 \pi )^3} \, e^{i p(x-y)} \, \epsilon_{\mu \nu \tau} {p^\tau \over p^2}  \; , \nonumber \\ 
 \WT{M^a (x)A}\! \null^b_\mu (y) &=&  \delta^{ab} \int {d^3p \over (2 \pi )^3} \, e^{i p(x-y)} \, {p_\mu \over p^2} \; , \\ 
 \WT{c^a (x) \, {\overline c}}\! \null^b(y) &=& 
i \, \delta^{ab} \int {d^3p \over (2 \pi )^3} \, e^{i p(x-y)} \, {1\over p^2} \; , \nonumber 
\label{2.11}
 \eea
where $p^\mu = \delta^{\mu \nu } p_\nu $ and $p^2 = \delta^{\mu \nu } p_\mu p_\nu$.  As it has been recalled in \cite{EG1,EF1}, the ratio (\ref{2.10}) of two functional integrations is understood as the limit ---in which the number of degrees of freedom tends to infinity--- of the ratio of two well defined standard integrals that are computed with a finite number of integration variables. 

\section{BRST symmetry}

The  gauge-fixed Chern-Simons theory in ${\mathbb R}^3$ is specified by the action $S_{TOT} = S_{CS} + S_{\phi \pi}$. Since only the gauge-fixing term $S_{\phi \pi }$   depends on the metric $g_{\mu \nu}$, the problem of establishing whether $Z$  necessarily nontrivially  depends on  $g_{\mu \nu}$  is really the problem of questioning the validity of the BRST gauge-fixing procedure. 

 In any quantum gauge theory, the introduction of a gauge-fixing procedure is not obligatory,  and does not enter the definition of the theory. 
 For instance, in the (finite) lattice regularization of gauge theories, by means of numerical calculus one can compute the expectation values of gauge-invariant functions of the fields without the necessity of using a gauge-fixing. So, the possible introduction of a gauge-fixing procedure does not enter the definition of the Chern-Simons theory. 

Suppose that, in the Chern-Simons field theory,  a unavoidable nontrivial dependence of $Z$ on the gauge-fixing metric $g_{\mu \nu}$ is demonstrated. In this case, this would not imply that the quantum Chern-Simons theory  depends on the framing of  3-manifold. Instead, this would mean that the  BRST gauge-fixing procedure is erroneous. Moreover,  the existence of a particular "preferred framing of 3-manifold" in which the  value of the overall phase factor becomes trivial does not mean that, in this ``particular framing", the path integration defines a topological invariant. In facts if, under a  modification of the trivialisation of the tangent bundle of 3-manifold, $Z$ gets modified, then the functional integration nontrivially depends on the gauge-fixing metric and it does not represent  a topological invariant of the type considered in low-dimensional topology  \cite{RO,PC}. 

There are two counterexamples showing that the claim that $Z$ depends on the gauge fining metric is false. 
The complete path integral solutions of the  $U(1)$ Chern-Simons and abelian {\rm BF}  theories have been produced in \cite{EF2,MT1,MT2} for  arbitrary closed oriented 3-manifolds. In these cases,   it has been shown that in the definition and in the computation of the functional  integration  there is no need of introducing a gauge-fixing term,  and therefore there is no need of introducing a metric in the 3-manifold. In the abelian Chern-Simons and BF theories,   the partition function has nothing to do with the perturbative gauge-fixing procedure. The partition function is a topological invariant which  is uniquely determined by the quadratic form \cite{FJ} on the torsion component of the homology group of the manifold. 

Let us recall that the Chern-Simons and BF expectation values 
of the Wilson line variables associated with oriented knots or links depend on the framing of the knots. This fact  does not contradict  the topological nature of the observables. Indeed, the framing of knots is a topological concept {\rm{\cite{RO}}}; no  metric enters the definition of the  trivialisation of a tubular neighbourhood of a knot  {\rm{\cite{RO}}}. The dependence of the Wilson line expectation values on the framing of the knots is due to  the definition of the holonomy as a well defined composite operator in quantum field theory. 

In perturbative quantum field theory, 
the validity of the gauge fixing procedure
is ensured by the BRST symmetry. 
As long as the BRST symmetry is an exact symmetry realized {\it  \`a la} Wigner,  the ground state is BRST invariant, and the expectation values of the gauge invariant observables are  gauge-fixing independent. 

Let us examine the possibility that, in the non abelian case, the BRST symmetry is invalidated by the presence of an anomaly. Let ${\cal V}$ be the set of the local  functionals of the type 
\be
{\cal V} = \left \{  \int d^3x \, c^a(x)  P^a (A^b_\nu (x) ) \right \} \; , 
\label{3.1} 
\ee
where $P^a (A^b_\nu (x) ) $ is a local polynomial  of the gauge fields $A^b_\nu (x)$ and their space-time derivatives. One can  verify that if  an element of $\cal V$ is BRST-closed,  
\be
\delta \left ( \int d^3x \, c^a(x)  P^a (A^b_\nu (x) )  \right ) =0 \; , 
\label{3.2} 
\ee 
then, in three dimensions,  this element is necessarily BRST-exact,  that is 
\be
 \int d^3x \, c^a(x)  P^a (A^b_\nu (x) ) = \delta \left ( \int d^3x \, R ((A^b_\nu (x) ) \right ) \; , 
 \label{3.3}
\ee
where $R ((A^b_\nu (x) )  $ is a local polynomial of the gauge fields (and their derivatives).  Consider now the perturbative construction of the renormalized effective action. If the ultraviolet divergences have been eliminated up to the $n$ loops level, the one-particle-irreducible diagrams with $(n+1)$ loops may contain divergences which are primitive divergences only. In this case, all the ambiguities (or divergences) of the corresponding amplitudes  are associated with local functions of the fields; for instance,  the primitive divergences can be eliminated by local counterterms.  Possible perturbative violations of a lagrangian symmetry are necessarily related with the ambiguities of the diagrams amplitudes. Therefore,  if there are no anomalies at the $n$ loops level,  the BRST variation of  the $(n+1)$ loops diagrams  with only primitive divergences   are elements of $\cal V$ which are also BRST-closed, because $\delta^2 =0$.   But elements  of this type are necessarily BRST-exact. Consequently, any nontrivial variation under BRST transformations  can be eliminated by means of a local counterterm. Thus,  at each finite order of the loop expansion, any nonvanishing primitive variation of the regularized  effective action can always be compensated iteratively by  local counterterms.  This means that, in the computation of the renormalized effective action,   it is possible to preserve the BRST symmetry at each order of perturbation theory.  
 This is precisely the reason why there are no gauge anomalies in three dimensions. 

\section{Sum of the diagrams}

Let us now compute the amplitudes associated with  all the Feynman diagrams contributing to the function  $\Phi ( g_{\mu \nu})$, which belongs to a possibly metric dependent   multiplicative factor  in the partition function 
\be
 Z = \frac{\int D ({\hbox{ fields }}) \, e^{i S_{TOT} }}{\int D ({\hbox{ fields }}) \, e^{i S_0 } } =   e^{i  \Phi ( g_{\mu \nu}) } \, Z_0 \; , 
 \label{4.1}
\ee
in which $Z_0$ does not depend on $g_{\mu \nu}$.  Let us consider the case, envisaged in \cite{WDG,AT},  in which  $\Phi (g_{\mu \nu})$ does not depend on the Chern-Simons  coupling constant $k$.  We need to concentrate only on the vacuum-to-vacuum diagrams which are connected, because   $\Phi ( g_{\mu \nu})$ appears in the exponential.  Since $\Phi (  g_{\mu \nu})$ does not depend on the Chern-Simons coupling constant, we must examine the diagrams   containing an arbitrary number of Wick contractions $ \WT{MA}$ and 
$ \WT{c \, {\overline c}}$, and in which the number of $\WT{A A}$  propagators is  equal to the number of $AAA$ interaction vertices. All these diagrams are one-loop diagrams constructed with the interaction vertices $h^{\mu \nu} \der_\mu M^a A^a_\nu$ and $h^{\mu \nu } \der_\mu  {\overline c}^a  \der_\nu c^a$. For instance, the contributions  which are linear in $h^{\mu \nu}$ are given by the sum of two terms  
\be
H_{1a} = i \int d^3x  \,  h^{\mu \nu} (x) \, 
 \WT{A^a_\nu (x) \, \der_\mu M} \! \null^a(x)  \; , 
\label{4.2}
\ee
and 
\be
H_{1b} = - i \int d^3x \,  h^{\mu \nu} (x) \, 
 \der_\nu \! \WT{c^a (x) \, \der_\mu {\overline c}}\! \null^a(x) \; . 
 \label{4.3}
\ee
The amplitudes $H_{1a}$ and $H_{1b}$ correspond to the diagrams shown in  Figure~1(a) and Figure~1(b) respectively, where the external wiggly line represents $h^{\mu \nu}$.

\vskip 0.5 truecm 
\centerline {
\begin{tikzpicture}[line width=1.0 pt, scale=1.2]
	\draw[vector] (0:1)--(0,0);
	\draw[postaction] (1,0) arc (180:0:.5);
	\draw[postaction] (2,0) arc (0:-180:.5);
	\begin{scope}[shift={(5,0)}]
	\draw[vector] (0:1)--(0,0);
	\draw[scalarnoarrow] (1,0) arc (180:0:0.5);
	\draw[scalarnoarrow] (2,0) arc (0:-180:0.5); 
	\end{scope}
		\node at (1.5,-1.2){(a)}; 
		\node at (6.5,-1.2){(b)};
	\end{tikzpicture}}
\vskip 0.3 truecm 
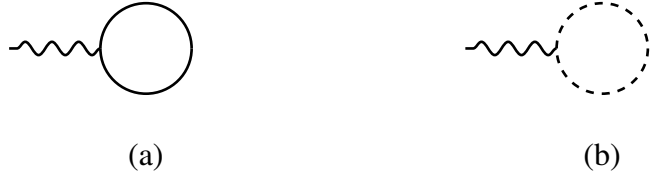
\centerline {{Figure~1.} ~Diagrams associated with contributions which are linear in $h^{\mu \nu}$.}
\vskip 0.3 truecm

\noindent Contributions to $\Phi $ which are quadratic functions of $h^{\mu \nu} $ are given by 
\be
H_{2a} = - \mezzo \int d^3x \, d^3y \, h^{\mu \nu } (x ) h^{\lambda \sigma} (y) \, 
\WT{A^b_\sigma (y) \, \der_\mu M} \! \null^a(x) 
\WT{A^a_\nu (x) \, \der_\lambda M} \! \null^b(y) \; , 
\label{4.4}
\ee
and
\be
H_{2b} = \mezzo \int d^3x \, d^3y \, h^{\mu \nu } (x ) h^{\lambda \sigma} (y) \, 
 \der_\sigma \! \WT{c^b (y) \, \der_\mu {\overline c}}\! \null^a(x) \, 
  \der_\nu \! \WT{c^a (x) \, \der_\lambda {\overline c}}\! \null^b(y) \; ,  
  \label{4.5}
\ee
which correspond to the diagrams of Figure~2(a) and Figure~2(b) respectively.

The generic contribution of order $n$ (with $n \geq 2$) in powers of $h^{\mu \nu}$ is the sum of two terms. The first term $H_{na}$ corresponds to a one-loop diagram, with  commuting variables running along the loop,  with $n$ vertices  of the type   $h^{\mu \nu} \der_\mu M^a A^a_\nu$,  
\bea
H_{na} &=& \frac{i^n}{n}\int d^3x_1 \cdots d^3x_n \, h^{\mu_1 \nu_1} (x_1)\cdots h^{\mu_n \nu_n}(x_n) \, \times \nonumber  \\
&& {\hskip 2 cm} \times \, \WT{A^{a_n}_{\nu_n} (x_n) \, \der_{\mu_1} M} \! \null^{a_1}(x_1) \cdots \WT{A^{a_{n-1}}_{\nu_{n-1}} (x_{n-1}) \, \der_{\mu_n} M} \! \null^{a_n}(x_n) \; . 
\label{4.6}
\eea

\vskip 0.6 truecm 
\centerline {
\begin{tikzpicture}[line width=1.0 pt, scale=1.2]
	\draw[vector] (0:1)--(0,0);
	\draw[postaction] (1,0) arc (180:0:.5);
	\draw[postaction] (2,0) arc (0:-180:.5);
	\draw[vector] (2,0)--(3,0);
	\begin{scope}[shift={(5,0)}]
	\draw[vector] (0:1)--(0,0);
	\draw[scalarnoarrow] (1,0) arc (180:0:0.5);
	\draw[scalarnoarrow] (2,0) arc (0:-180:0.5); 
	\draw[vector] (2,0)--(3,0);
	\end{scope}
		\node at (1.5,-1.2){(a)}; 
		\node at (6.5,-1.2){(b)};
	\end{tikzpicture}}
\vskip 0.3 truecm 
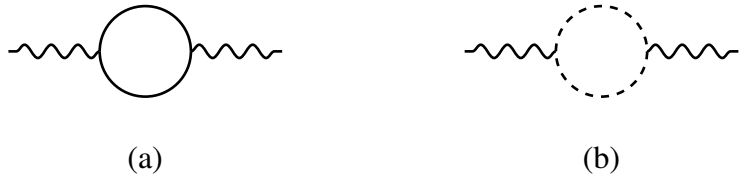
\centerline {{Figure~2.} ~Contributions to $\Phi $ which are quadratic in $h^{\mu \nu}$.}
\vskip 0.3 truecm

The second term $H_{nb}$ is associated with a one-loop diagram,  with ghost/antighost  fields running along the loop,  with  $n$ vertices of the type $h^{\mu \nu } \der_\mu  {\overline c}^a  \der_\nu c^a$, 
\bea
H_{nb} &=& - \frac{i^n}{n}\int d^3x_1 \cdots d^3x_n \, h^{\mu_1 \nu_1} (x_1)\cdots h^{\mu_n \nu_n}(x_n) \, \times \nonumber  \\
&& {\hskip 1.7 cm} \times \, 
 \der_{\nu_n} \! \WT{c^{a_n} (x_n) \, \der_{\mu_1} {\overline c}}\! \null^{a_1}(x_1)  
 \cdots  \der_{\nu_{n-1}} \! \WT{c^{a_{n-1}} (x_{n-1}) \, \der_{\mu_n} {\overline c}}\! \null^{a_n}(x_n)  \; . 
\label{4.7}
\eea
Note that, with respect to the $H_{na}$ expression shown in equation (\ref{4.6}),  an additional minus sign appears in front of $H_{nb}$; this is due to the rearrangement of the anti-commuting ghost fields in the propagators. As in quantum electrodynamics, each loop of anti-commuting fields  brings  an extra minus sign \cite{IZ}. The sum of all the contributions gives 
\be
i \,\Phi (g_{\mu \nu}) = \sum_{n=1}^\infty       \left (    H_{na} + H_{nb}     \right ) \; . 
\label{4.8}
\ee
By using the field propagators shown in equation (2.11), one finds that all terms $ H_{na}$ and  $ H_{nb}  $ have ultraviolet divergences. So we must  introduce a regularization. At the regularized level, the validity of the gauge-fixing procedure requires a regularisation that preserves the BRST symmetry.  Note that the BRST variation of the operator $A^a_\mu(x) \der_\nu {\overline c}^b (y)$ is given by 
\be
\delta \left ( A^a_\mu(x) \der_\nu {\overline c}^b (y) \right ) = \der_\mu c^a(x) \der_\nu {\overline c}^b (y) - f^{abc} A_\mu^b (x)c^c(x) \der_\nu {\overline c}^b (y) - A^a_\mu(x) \der_\nu M^b (y) \; . 
\label{4.9}
\ee
Let us now consider the vacuum expectation values of the  field operators entering equation (\ref{4.9}). 
When the ground state is BRST invariant, the expectation value of $\delta \left ( A^a_\mu(x) \der_\nu {\overline c}^b (y) \right ) $ is vanishing.  Indeed, the BRST variation of a generic  operator $O_p$ can be written as $\delta O_p = \{ Q , O_p\} $ where $Q$ denotes the generator of the BRST transformations. If the ground state $| \Omega \rangle $ is BRST invariant, it satisfies $Q | \Omega \rangle = 0 $. Therefore, $\langle \Omega | \delta O_p | \Omega \rangle = \langle \Omega | \{ Q , O_p \} | \Omega \rangle =0 $. 
The computation of the expectation values of the operators appearing on the right-hand-side of equation (\ref{4.9}) admits a perturbative expansion. At lowest order in perturbation theory ---corresponding to the diagrams with no "interaction vertices"---   from equation (\ref{4.9}) one finds 
\be
0 =  \der_\mu \! \WT{c^a (x) \, \der_\nu {\overline c}}\! \null^b(y) - 
\WT{A^a_\mu (x) \, \der_\nu M} \! \null^b(y) \; . 
\label{4.10}
\ee
Equation (\ref{4.10}) is indeed satisfied by the 
propagators shown in equation (2.11).
 What's more, the argument that we have presented to  derive  equation (\ref{4.10}) shows that, in order to maintain the BRST invariance at the regularized level,  the regularized propagators also must satisfy relation (\ref{4.10}), 
\be
\WT{A^a_\mu (x) \, \der_\nu M} \! \null^b(y)  \biggr |_{reg} = \der_\mu \! \WT{c^a (x) \, \der_\nu {\overline c}}\! \null^b(y) \biggr |_{reg} \; . 
\label{4.11}
\ee
One regularisation of this type is given by 
\be
\WT{A^a_\mu (x) \, \der_\nu M} \! \null^b(y)  \biggr |_{reg} = \der_\mu \! \WT{c^a (x) \, \der_\nu {\overline c}}\! \null^b(y) \biggr |_{reg} = i \, \delta^{ab} \int {d^3p \over (2 \pi )^3} \, e^{i p(x-y)} \, {p_\mu p_\nu \over p^2} \, e^{- \epsilon p^2} \; , 
\label{4.12}
\ee
where $\epsilon > 0 $ represents a regularisation parameter. The regularisation is removed in the $\epsilon \rightarrow 0 $ limit.    
By using the regularized propagators (\ref{4.12}), all  amplitudes $H_{na}$ ad $H_{nb}$ are well defined and finite. Since $H_{na} = - H_{nb}$, equation (\ref{4.8}) shows  that 
\be
\Phi (g_{\mu \nu}) =0 \; . 
\label{4.13}
\ee   
The direct computation shows that, in the partition function,  the presence of a  multiplicative factor which nontrivially depends on the gauge-fixing metric is excluded. 

An immediate generalisation of the result (\ref{4.13}) concerns the expectation values of the Wilson line operators. Let $W(L)$ be the product of the traces  of the gauge holonomies (in arbitrary representations) associated with the components of an oriented link $L \subset {\mathbb R}^3$.  Also in this case, one could conjecture that the expectation value of $W(L)$ contains a multiplicative factor which nontrivially depends on the gauge-fixing metric, 
\be
 Z(L) = N^{-1} \int D ({\hbox{ fields }}) \, e^{i S_{TOT} } \, W(L) =   e^{i  \Phi^\prime ( g_{\mu \nu}) } \, Z_0(L) \; , 
 \label{4.14}
\ee
in which $Z_0(L)$ does not depend on $g_{\mu \nu}$. Let us examine the case in which $ \Phi^\prime ( g_{\mu \nu}) $ does not depend on the Chern-Simons coupling constant.  In these circumstances, the vacuum-to-vacuum Feynman diagrams contributing to $ \Phi^\prime ( g_{\mu \nu}) $ are exactly  the same diagrams  contributing to 
$ \Phi ( g_{\mu \nu}) $, that have been considered above. Therefore, also in this case  one finds  
\be
\Phi^\prime (g_{\mu \nu}) =0 \; . 
\label{4.15}
\ee   

To sum up, the nonabelian  Chern-Simons field theory  in ${\mathbb R}^3$ admits a canonical quantization and, because of  the exact BRST symmetry, even in the presence of gauge-fixing lagrangian terms which contain a gauge-fixing metric, the partition function  and the expectation values of the Wilson line operators do not depend on the gauge-fixing metric through a multiplicative factor which does not depend on the Chern-Simons coupling constant.   

\section{BF model}

The action $\widetilde S$  of the BF model in ${\mathbb R}^3$, with gauge group $ISU(2)$,    is given by \cite{GR}
\be
\widetilde S = \int d^3 x \, \epsilon^{\mu \nu \lambda}  \left \{ \mezzo  \, B^a_\mu F^a_{\nu \lambda} (A) + g  \left [ \mezzo A^a_\mu \der_\nu A^a_\lambda
 - \sesto \epsilon^{abc }\, A^a_\mu A^b_\nu A^c_\lambda \right ] \right \} \; , 
 \label{5.1}
 \ee
where $A^a_\mu (x)$ and $B^a_\mu (x)$ (with $a=1,2,3$) are the local components of the $ISU(2)$ connection, and $g$ is a real coupling constant.  The BRST transformations \cite{WA,GR} are given by 
\bea
\delta A_\mu^a =\der_\mu c^a - \epsilon^{abc} A_\mu^b c^c  \; &,& \; \delta B^a_\mu = \der_\mu \xi^a - \epsilon^{abc} A_\mu^b \xi^c
- \epsilon^{abc} B_\mu^b c^c \; , \nonumber \\
\delta c^a = \mezzo \epsilon^{abc} c^bc^c \quad  
, \quad \delta {\overline c}^a = - M^a \; &,& \; 
\delta \xi^a = \epsilon^{abc} \xi^b c^c   \quad , \quad \delta {\overline \xi}^a = - N^a \; ,  \\   \delta M^a =0 \; &,& \; \delta N^a =0 \; , \nonumber 
\label{5.2}  
\eea
where $\{  \xi^a , \overline \xi^a, c^a , {\overline c}^a   \}$ is the set of anticommuting ghosts and antighosts fields,  whereas $M^a , N^a $ represent the commuting auxiliary  fields.  In the Landau gauge,  the gauge-fixing and ghosts action terms are given by 
\bea
\widetilde S_{\phi \pi } &=& \lambda \int d^3x  \, \sqrt g \, g^{\mu \nu } \, \Bigl \{ \der_\mu M^a A^a_\nu + \der_\mu N^a  B^a_\nu + \der_\mu  {\overline c}^a  (\der_\nu c^a - \epsilon^{abc} A_\nu^b c^c ) \nonumber \\ 
&& {\hskip 1.4 cm} + \der_\mu  {\overline \xi}^a  (\der_\nu \xi^a - \epsilon^{abc} A_\nu^b \xi^c
- \epsilon^{abc} B_\nu^b c^c) \Bigr \}  \nonumber \\ 
&& = \delta \left [ \lambda \int d^3x  \, \sqrt g \, g^{\mu \nu } \, 
 \Bigl \{  - \der_\mu  {\overline c}^a  A^a_\nu - \der_\mu  {\overline \xi}^a B^a_\nu \Bigr \} \right ]  \; , 
\label{5.3} 
\eea
where $g_{\mu \nu} $ represents the gauge-fixing metric and $\lambda \not= 0$ is a real free parameter. Also in this case, by means of the definitions (\ref{2.5}) and (\ref{2.6}) we can decompose the total action $\widetilde S_{TOT}$  into a sum of two terms 
 \be
\widetilde S_{CS} + \widetilde S_{\phi \pi} = \widetilde S_{TOT} = \widetilde S_0 + \widetilde S_I \; , 
\label{5.4}
 \ee
where $\widetilde S_0$ does not depend on $g_{\mu \nu}$,  
\bea
 \widetilde S_0 &=& \int d^3 x  \Bigl [ \epsilon^{\mu \nu \rho} B_\mu^a  \der_\nu A^a_\rho +  \gmezzi  \epsilon^{\mu \nu \rho}   A^a_\mu \der_\nu A^a_\rho +  \delta^{\mu \nu }\der_\mu M^a  A^a_\nu \nonumber \\ 
 && {\hskip 1.6 cm} + \delta^{\mu \nu }\der_\mu N^a  B^a_\nu  +  \delta^{\mu \nu }  \der_\mu  {\overline c}^a  \der_\nu c^a  + 
  \delta^{\mu \nu }  \der_\mu  {\overline \xi}^a  \der_\nu \xi^a
 \Bigr ] \; , 
 \label{5.5}
 \eea
and 
\bea
\widetilde S_I &=& \int d^3x \Bigl [  h^{\mu \nu} \der_\mu M^a A^a_\nu +  h^{\mu \nu} \der_\mu N^a  B^a_\nu + h^{\mu \nu} 
 \der_\mu  {\overline c}^a  \der_\nu c^a + h^{\mu \nu} \der_\mu  {\overline \xi}^a  \der_\nu \xi^a \nonumber \\
&& \qquad  \qquad  - \mezzo \epsilon^{\mu \nu \rho} \epsilon^{abc}B^a_\mu A^b_\nu A^c_\rho  -   \gsesti \epsilon^{\mu \nu \rho}  \epsilon^{abc }\, A^a_\mu A^b_\nu A^c_\rho 
- \lambda  \sqrt g \, g^{\mu \nu } \epsilon^{abc} \der_\mu  {\overline c}^a A_\nu^b c^c \nonumber \\ 
&& \qquad \qquad 
- \lambda  \sqrt g \, g^{\mu \nu } \epsilon^{abc} \der_\mu  {\overline \xi}^a \left (   A_\nu^b \xi^c + B^b_\nu c^c          \right )
\Bigr ] \; . 
\label{5.6}
\eea
The nontrivial components of the fields propagators which are specified by the ``free" action $\widetilde S_0$ are given by 
\bea
\WT{A^a_\mu (x) B}\! \null^b_\nu(y) &=& \delta^{ab} \int {d^3p  \over (2 \pi )^3} \, e^{i p(x-y)} \, \epsilon_{\mu \nu \lambda} {p^\lambda \over p^2}   \; , \nonumber \\ 
\WT{B^a_\mu (x) B}\! \null^b_\nu(y) &=& - g \, \delta^{ab} \int {d^3p \over (2 \pi )^3} \, e^{i p(x-y)} \, \epsilon_{\mu \nu \lambda} {p^\lambda \over p^2}  \; , 
\label{5.7}
\eea
\bea
\WT{M^a (x) A}\! \null^b_\mu (y) &=&  \delta^{ab} \int {d^3p \over (2 \pi )^3} \, e^{i p(x-y)} \, {p_\mu \over p^2} \; , \nonumber \\ 
\WT{N^a (x) B}\! \null^b_\mu (y) &=&  \delta^{ab} \int {d^3p \over (2 \pi )^3} \, e^{i p(x-y)} \, {p_\mu \over p^2}   \; , 
\label{5.8}
\eea
and  
\be
\WT{c^a (x) \, {\overline c}}\! \null^b(y) = 
\WT{\xi^a (x) \, {\overline \xi}}\! \null^b(y) = 
i \delta^{ab} \int {d^3p \over (2 \pi )^3} \, e^{i p(x-y)} \, {1\over p^2} \; . 
\label{5.9}
\ee

The normalized partition function $\widetilde Z$ of the BF model is defined as 
\be
 \widetilde Z = \frac{\int D ({\hbox{ fields }}) \, e^{i \widetilde S_{TOT} }}{\int D ({\hbox{ fields }}) \, e^{i \widetilde S_0 } } 
 = e^{i \widetilde \Phi (g_{\mu \nu})} \widetilde Z_0 \; , 
 \label{5.10}
\ee
where $\widetilde Z_0 $ does not depend on $g_{\mu \nu}$ and $\widetilde \Phi (g_{\mu \nu})$ denotes a possibly nontrivial factor depending on the gauge-fixing metric $g_{\mu \nu}$.  The connected vacuum-to-vacuum diagrams contributing to $\widetilde \Phi (g_{\mu \nu})$ admit an expansion in powers of the coupling constant $g$. The lowest order contributions correspond to one-loop diagrams constructed with the vertices belonging to $\widetilde S_I$ which are quadratic in the field operators.  The diagrams containing the vertices $h^{\mu \nu} \der_\mu M^a A^a_\nu$  and  the diagrams containing the vertices $ h^{\mu \nu} \der_\mu N^a  B^a_\nu $ correspond to  amplitudes which are equal;  these amplitudes coincide with the ``bosonic" amplitudes  of the Chern-Simons theory. Similarly,  the diagrams containing the vertices 
 $ h^{\mu \nu}  \der_\mu  {\overline c}^a  \der_\nu c^a $ and the diagrams containing the vertices  $ h^{\mu \nu} \der_\mu  {\overline \xi}^a  \der_\nu \xi^a $ are associated with the same amplitudes that are equal to the ``ghost" amplitudes of the Chern-Simons theory. Therefore, the BF diagrams correspond  to twice the diagrams of the Chern-Simons theory;  this is in agreement with the general property \cite{GR} that, with appropriate external sources,  the one loop generating functional of the BF theory is twice the generating functional of the Chern-Simons theory. So, by means of a BRST invariant regularization of the type shown in equation (\ref{4.12}), one finds that also in the BF model the metric dependent phase factor  $\widetilde \Phi (g_{\mu \nu})$ is vanishing.    

\section{Independence of the gauge-fixing metric}

In this section  we present a rather general argument showing that, for renormalizable gauge theories in three and four dimensions, the partition function and the mean values of the gauge invariant  observables do not depend on the gauge-fixing metric.  

Let us consider a  gauge invariant theory specified by the action $S$, which depends on the fields $A^a_\mu (x)$  of the gauge connection and possibly on  matter fields.  We assume that $S$ is invariant under local $SU(N)$ gauge transformations; the usual infinitesimal  gauge transformations $\Delta A_\mu^a (x)$ of  the gauge fields are given by 
\be
\Delta A_\mu^a (x) = \der_\mu \theta^a(x) - f^{abc} A_\mu^b (x) \theta^c (x) \; , 
\label{6.1}
\ee
where $\theta^a(x)$ are the infinitesimal gauge parameters. 
The matter fields are organized in $SU(N)$ multiplets. We have already mentioned the fact that in three space-time  dimensions there are no gauge anomalies; so, in three dimensions we only assume standard renormalizability.  In four dimensions, 
the renormalizability also requires the absence of  Adler-Bardeen anomalies \cite{ADL,BAR,BEL,HBN}.  For instance, when matter fields describe spin 1/2 fermions, we shall require that 
 the sum over the matter multiplets of the symmetric function 
 ${\rm Tr}  \left ( R^a \{ R^b , R^c \} \right )$
 of the $SU(N)$ generators  $\{ R^a \}$ is vanishing 
\be
\sum_{\rm matter ~ multiplets} {\rm Tr} \left ( R^a \{ R^b , R^c \} \right ) =0 \; . 
\label{6.2}
\ee
When condition (\ref{6.2}) is satisfied, the matter content of the theory is anomaly free, {\it i.e.} there are no Adler-Bardeen gauge anomalies \cite{AB,ZU,BZ,MSZ}.  The BRST transformations act on the matter fields as infinitesimal local gauge transformations in which the gauge parameter $\theta^a(x)$ must be replaced by the ghost field $c^a(x)$. For the gauge fields and the remaining fields, the BRST transformations are determined by the structure of the gauge group 
  \bea
\delta A_\mu^a =\der_\mu c^a - f^{abc} A_\mu^b c^c  \; &,& \;  \delta c^a = \mezzo f^{abc} c^bc^c \; , \nonumber \\
 \delta {\overline c}^a = - M^a \; &,& \; 
 \delta M^a =0 \; .   
 \label{6.3}
\eea
The gauge-fixing part of the action, corresponding to the Landau gauge, is given by   
\be
S_{\phi \pi }  =  \lambda \int d^nx \, \sqrt g \, g^{\mu \nu } \, \Bigl \{ \der_\mu M^a  A^a_\nu +  \der_\mu  {\overline c}^a  (\der_\nu c^a - f^{abc} A_\nu^b c^c )  
 \Bigr \} \; ,  
 \label{6.4}
\ee
where $\lambda $ is a free real parameter,  the integer $n$ denotes the dimensions ($n=3$ or $n=4$), 
 and the gauge-fixing metric $g_{\mu \nu}$ is a generic metric in three or four dimensions, which a priori is not related with the ``physical metric" which possibly appears in $S$.  Expression (\ref{6.4}) can be written as 
\be
S_{\phi \pi } = \lambda \, \delta \Psi \; , 
\label{6.5}
\ee
where 
\be
\Psi =  -  \int d^nx \, \sqrt g \, g^{\mu \nu } \,   \der_\mu  {\overline c}^a  \, A^a_\nu  \; . 
 \label{6.6}
\ee
Let $X$ be a gauge invariant function of the gauge fields $A_\mu^a (x) $ and of the matter fields. Since $X$ is invariant under gauge transformations, $X$ is also invariant under BRST transformations. Let us consider  the mean value 
\be
Z  (X) = N^{-1} \int D ({\hbox{ fields }}) \, e^{i S + i S_{\phi \pi}} \, X \; , 
 \label{6.7}
\ee
  where the normalization factor $N$ does not depend on the gauge-fixing metric $g_{\mu \nu}$ and does not depend on $\lambda$. As it is shown in equation (\ref{6.4}),  the parameter $\lambda $ only appears in $S_{\phi \pi}$; therefore
\be
- i \, \frac{\der Z(X)}{\der \lambda} = N^{-1}  \int D ({\hbox{ fields }}) \, e^{i S + i S_{\phi \pi}} \, X \,  \delta \Psi \; . 
\label{6.8}
\ee
Since $X$ is BRST invariant, one has $X \, \delta \Psi  = \delta \left ( X \, \Psi \right )$, and then 
\be
- i \, \frac{\der Z(X)}{\der \lambda} = N^{-1}  \int D ({\hbox{ fields }}) \, e^{i S + i S_{\phi \pi}} \, \delta \left (  X \,   \Psi \right ) = 0  \; . 
\label{6.9}
\ee
The last equality in equation (\ref{6.9}) is a consequence of the exact BRST symmetry. Indeed, on the one side, if the ground state is BRST invariant,  the mean value of a BRST variation is vanishing.  On the other side, since the total action is BRST invariant, a change of variables in the functional integration shows that the mean value of a BRST variation is vanishing. 
The validity of both arguments is guaranteed, order by order in perturbation theory, by the renormalizability of the theory and by the absence of gauge anomalies. Since the gauge-fixing metric and $\lambda $ only appear in $S_{\phi \pi}$, and since  $S_{\phi \pi}$ is a linear function of $\lambda $, equation (\ref{6.9}) shows that $Z(X)$ does not depend on the gauge-fixing metric.  

\section{Conclusions}

Topological gauge theories in three dimensions are characterized by metric independent lagrangians. As a consequence of the introduction of a gauge-fixing term which depends on a gauge-fixing metric $g_{\mu \nu}$, the total gauge-fixed lagrangian depends on the gauge-fixing metric. However, in agreement with the general properties of quantum field theories, one expects that 
the partition function and the mean values of the gauge invariant variables do no depend on $g_{\mu \nu}$.  

This is confirmed by the complete solution of the abelian Chern-Simons and BF theories defined in a generic closed 3-manifold.  

In the present article we have considered the nonabelian Chern-Simons and the BF theories  in ${\mathbb R}^3$, where a canonical quantization exists and the perturbative expansion can be defined with the help of the BRST gauge-fixing procedure. We have shown that, since there are no gauge anomalies in three dimensions, the BRST symmetry guarantees the complete metric independence of the observables.  The nontrivial dependence of the partition function on the gauge-fixing metric by means of an overall phase factor ---which does not depend on the coupling constant of the theory---  has been excluded by direct computation. 
Finally, a rather general argument has been presented showing that, for renormalizable gauge theories in three dimensions and in four dimensions, the partition function and the mean vaules of the gauge invariant observables do not depend on the gauge-fixing metric.  

\bigskip

\centerline {\bf Added Notes}

\medskip 

In the present article we have considered the BRST symmetry of  topological field theories within the perturbative approach. Let us add some final notes on  possible non-perturbative effects which could break the BRST symmetry. Let us remind that, in the absence of gauge anomalies,  the BRST symmetry can be preserved \cite{PSB} at each finite order of the loop expansion.  In the BF model, the effective action only contains 0-loop and 1-loop proper vertices \cite{GR}; therefore, non-perturbative effects  are absent. 
The nontrivial dependence of the Chern-Simons partition function on the gauge-fixing metric ---through an overall phase factor which does not depend on the coupling constant--- is 
a perturbative issue. Indeed, as we have shown in Section~4, 
this phase factor is entirely determined by the amplitudes associated with one-loop Feynman diagrams.  
It should be noted that, in the abelian case with gauge group $U(1)$,  both Chern-Simons and BF models can also be solved by means of the Deligne-Beilinson formalism \cite{EF2,MT1,MT2} without the  introduction of any gauge fixing, and the exact solution coincides with the results of the perturbative approach. 
Since the 1-loop computations of Section~4  are valid in both  the abelian and the non-abelian cases, this provides an  additional proof that the functional integration does not depend on the gauge-fixing metric.  
Finally, in the Chern-Simons theory in ${\mathbb R}^3$, the perturbative computations are in complete agreement with the mathematical construction of Reshetikhin-Turaev \cite{RTP} by means of quantum groups; the mean values of the Wilson lines associated with links take the form of link polynomials that are analytic functions of $1/k$.   In this sense,  perturbation theory provides the exact solution for the expectation values of the Chern-Simons observables, and then potential non-perturbative effects  are excluded.


\vskip 1 truecm 

\begin{thebibliography}{AAA}

\bibitem{WI} E.~Witten, Commun. Math. Phys. 121 (1989) 351. 

\bibitem{HO} G.T.~Horowitz, Commun. Math. Phys. 125 (1989) 417.

\bibitem{KR} A.~Karlhede and M.~Rocek, Phs. Lett. B 224  (1989) 58.

\bibitem{MP} R.~Meyers and V.~Perival, Phys. Lett. B 225 (1989) 352.

\bibitem{BT}  M.~Blau and G.~Thompson,  
Phys. Lett. B 228  (1989) 64. 

 \bibitem{GR} E.~Guadagnini and F.~Rottoli, Nucl. Phys. B 954 (2020) 114987.  

\bibitem{WDG} D.~Bar-Natan and E.~Witten, Commun. Math. Phys. 141 (1991) 423. 

\bibitem{AT} M.~Atiyah, Topology, 29 (1990) 0040-9383/90 503.00 +.00. 


\bibitem{N} N.K.~Nielsen, Nucl. Phys. B 101 (1975) 173. 

\bibitem{KSJB} H.~Kluberg-Stern and J.B.~Zuber, Phys. Rev. D 12 (1975) 467. 

\bibitem{BLE} S.D.~Joglekar and B.W.~Lee, Ann. Phys. 97 (1976) 160. 

\bibitem{AFF} I.J.R.~ Aitchison and C.M.~Fraser, Ann. Phys. 156 (1984). 

\bibitem{SIB} O.~Piguet and K.~Sibold, Nucl. Phys. B 253 (1985) 517. 

\bibitem{HHK} R.~Haussling and E.~Kraus, Z. Phys. C 75 (1997) 739. 

\bibitem{AGM} A.~Aste, G.~Scharf and M.~D\"utsch, J. Phys. A: Math. gen. 31 (1998) 1563. 

\bibitem{DCI} O.M.~ Del Cima, Phys. Lett. B 457 (1999) 307. 

 
 \bibitem{KAU} R.K.~Kaul and R.~Rajaraman, Phys. Lett. B 249 (1990) 433. 
 
  \bibitem{CUG} L.F.~Cigliandolo, G.~Lozano, F.A.~Schaposnik, Phys. Lett. B 253 (1991) 90.
 
 \bibitem{MAT} M.~Blau and G.~Thompson, Phys. Lett. B 255 (1991) 535. 
 
  
 \bibitem{DAN} D.~Birmingham, R.~Gibbs and S.~Mokhtari, Phys. Lett. B 273  (1991) 67.
 
 \bibitem{BRO} B.~Broda, Phys. Lett. B 280 (1992) 47. 
 
 \bibitem{ABU} M.~Abud and G.~Fiore, Phys. Lett. B 293 (1992) 89.
 
\bibitem{RES}  M.I.~Caicedo and A.~Restuccia, Phys. Lett. B 307 (1993) 77. 
 
  \bibitem{CAIC} M.I.~Caicedo, R.~Gianvittorio, A.~Restuccia, J.~Stephany, Phys. Lett. B 354 (1995) 292.  
  
  \bibitem{MTA} M.~Tahiri, Inter. J. of Mod. Phys. A 12 (1997) 3153.
 
 \bibitem{CHE} W.F.~Chen, H.C.~Lee and Z.Y.~Zhu, Phys. Rev. D 56 (1997) 1170. 
 
\bibitem{QUI} A.G.~Quinto and A.F.~Ferrari, Phys. Rev. D 94 (2016) 085006.

 
\bibitem{FO} V. A. Fock, Sov. Phys. 12, 404 (1937).

\bibitem{SC} J. Schwinger, Phys. Rev. 82, 684 (1952).

\bibitem{GAL} L.~Gallot, E.~Guadagnini, E.~Pilon and F.~Thuillier, Mod. Phys. Lett. A 29 (2014) 1450121. 

\bibitem{FR} L.~Gallot, E.~Pilon and F.~Thuillier, Mod. Phys. Lett. A 30 (2015) 1550102. 

\bibitem{BRS} C.~Becchi, A.~Rouet and R.~Stora, Commun. Math. Phys. 42 (1975) 127.   

\bibitem{2S} C.~Becchi, A.~Rouet and R.~Stora, Ann. Phys. (NY) 98 (1976) 287.

\bibitem{TYU} I.V.~Tyutin, Lebedev Institute preprint N39 (1975). 

\bibitem{BO} N.N.~Bogoliubov and D.V.~Shirkov, {\it Introduction to the theory of quantized fields}, John Wiley \& Sons (New York, 1980).

\bibitem{IZ} C.~Itzykson and J.-B.~Zuber, {\it Quantum Field Theory}, (McGraw-Hill,1980).


\bibitem{PS} M.E.~Peskin and D.V.~Schroeder, {\it An Introduction to Quantum Field Theory}, Westview Press (Boulder, 1995).

\bibitem{EG1} E.~Guadagnini, {\it Functional integration and abelian link invariants}, published in "Chern-Simons theory: 20 years after"  (AMS/IP Studies in Advanced Mathematics Volume 50, 2011) edited by J.E.  Andersen, H. Boden, A. Hahn and B. Himpel (Providence, RI: American Mathematical Society/International Press), ISBN-10: 0-8218-5353-8. 

\bibitem{EF1} E.~Guadagnini and F.~Thuillier, J. Math. Phys. 54  (2013) 082302.  

\bibitem{RO} D.~Rolfsen, {\it Knots and Links}, AMS Chelsea Publishing, Providence, 2003.

\bibitem{PC} P.~Cromwell, {\it Knots and links}, Cambridge University Press, (Cambridge, 2004). 
 
 \bibitem{EF2} E.~Guadagnini and F.~Thuillier,     Nucl. Phys. B 882 (2014) 450. 
 
 \bibitem{MT1} Ph.~Mathieu and F.~Thuillier, J. Math. Phys. 57 (2016), 022306.
 
\bibitem{MT2} Ph.~Mathieu, F.~Thuillier, J. Math. Phys. 58 (2017) 102301.
 
 \bibitem{FJ} {\it Topology of 3-manifolds, and related topics}, edited by M.K.~Fort, Jr.,   Dover Publications, INC. (New York, 2010). 
 
 \bibitem{WA} J.C.~Wallet, Phys. Lett. B 235  (1990) 71. 

\bibitem{ADL} S.L. Adler, Phys. Rev. 177 (1969) 2426. 

\bibitem{BAR} W.A. Bardeen, Phys. Rev. 184 (1969) 1848.

\bibitem{BEL}  J.S. Bell and R. Jackiw,  Nuovo Cim. A 60
(1969) 47.

\bibitem{HBN}  H.B. Nielsen and M. Ninomiya,  Phys. Lett. B 130 (1983) 389. 

 \bibitem{AB} S.L. Adler and W.A. Bardeen,  Phys. Rev. 182 (1969) 1517. 

\bibitem{ZU} B. Zumino, Y.-S. Wu and A. Zee,  Nucl. Phys. B 239 (1984) 477. 

\bibitem{BZ}  W.A. Bardeen and B. Zumino,  Nucl. Phys. B 244 (1984) 421.  

\bibitem{MSZ} J. Manes, R. Stora and B. Zumino,  Commun. Math. Phys. 102 (1985) 157.

\bibitem{PSB}  O. Piguet and S.P. Sorella, {\it Algebraic Renormalization}, Springer-Verlag (Berlin, 1995). 

\bibitem{RTP} N.Y.~Reshetikhin and V.G.~Turaev, Commun. Math. Phys. 127 (1990) 1. 

\end{thebibliography}
\end{document}